# Non-monotonic band flattening near the magic angle of twisted bilayer MoTe$_2$

Yujun Deng,[1,*] William Holtzmann,[2] Ziyan Zhu,[1,3] Timothy Zaklama,[4] Paulina Majchrzak,[5] Takashi Taniguchi,[6] Kenji Watanabe,[7] Makoto Hashimoto,[8] Donghui Lu,[8] Chris Jozwiak,[9] Aaron Bostwick,[9] Eli Rotenberg,[9] Liang Fu,[4] Thomas P. Devereaux,[1,10] Xiaodong Xu,[2,11] and Zhi-Xun Shen[1,5,12,13,†]

[1]*Stanford Institute for Materials and Energy Sciences, SLAC National Accelerator Laboratory, Menlo Park, CA 94025, USA*
[2]*Department of Physics, University of Washington, Seattle, WA 98195, USA*
[3]*Department of Physics, Boston College, Chestnut Hill, MA 02467, USA*
[4] *Department of Physics, Massachusetts Institute of Technology, Cambridge, MA 02139, USA*
[5] *Department of Applied Physics, Stanford University, Stanford, CA 94305, USA*
[6]*Research Center for Materials Nanoarchitectonics, National Institute for Materials Science, 1-1 Namiki, Tsukuba 305-0044, Japan*
[7]*Research Center for Electronic and Optical Materials, National Institute for Materials Science, 1-1 Namiki, Tsukuba 305-0044, Japan*
[8]*Stanford Synchrotron Radiation Lightsource, SLAC National Accelerator Laboratory, Menlo Park, CA 94025, USA*
[9]*Advanced Light Source, Lawrence Berkeley National Laboratory, Berkeley, CA 94720, USA*
[10]*Department of Materials Science and Engineering, Stanford University, Stanford, CA 94305, USA*
[11]*Department of Materials Science and Engineering, University of Washington, Seattle, WA 98195, USA*
[12]*Department of Physics, Stanford University, Stanford, CA 94305, USA*
[13]*Geballe Laboratory for Advanced Materials, Stanford University, Stanford 94305, CA, USA*

Twisted bilayer MoTe$_2$ (tMoTe$_2$) is an emergent platform for exploring exotic quantum phases driven by the interplay between nontrivial band topology and strong electron correlations. Direct experimental access to its momentum-resolved electronic structure is essential for uncovering the microscopic origins of the correlated topological phases therein. Here, we report angle-resolved photoemission spectroscopy (ARPES) measurements of tMoTe$_2$, revealing pronounced twist-angle-dependent band reconstruction shaped by orbital character, interlayer coupling, and moiré potential modulation. Density functional theory (DFT) captures the qualitative evolution, yet underestimates key energy scales across twist angles, highlighting the importance of electronic correlations. Notably, the hole effective mass at the K point exhibits a non-monotonic dependence on twist angle, peaking near 2°, consistent with band flattening at the magic angle predicted by continuum models. Via electrostatic gating and surface dosing, we further visualize the evolution of electronic structure versus doping, enabling direct observation of the conduction band minimum and confirm tMoTe$_2$ as a direct band gap semiconductor. These results establish a spectroscopic foundation for modeling and engineering emergent quantum phases in this moiré platform.

## I. INTRODUCTION.

Twisted van der Waals heterostructures have emerged as a powerful platform for engineering novel quantum phases of matter [1–4]. Stacking two-dimensional materials with controlled rotational misalignment results in moiré superlattices that profoundly alter the electronic structure, enabling correlated insulators, superconductivity, and a variety of topological phases, such as quantum anomalous Hall (QAH), fractional QAH (FQAH), and fractional quantum spin Hall (FQSH) states [5–12]. Among these systems, tMoTe$_2$ has attracted particular interest due to its experimental realization of both integer and fractional topological phases [10,11,13–18].

To date, experimental studies of tMoTe$_2$ have predominantly focused on transport measurements [10,11,14,17,18], unveiling a rich phase diagram featuring the aforementioned states. However, direct momentum-resolved probes of its electronic structure are crucial for a deep understanding of the underlying mechanisms. Theoretical studies based on DFT [19,20] and continuum models [20–23] predict significant twist-angle-dependent band reconstructions driven by the modulation of interlayer tunneling and moiré potential. Yet, key experimental benchmarks to validate these predictions have been lacking.

Here, we employ nanoARPES to systematically investigate the twist-angle-dependent electronic band structure of tMoTe$_2$. The obtained band structure information enables extraction of key band structure parameters including intra-valley band splittings, the energy difference between the valence band maxima (VBM) at Γ and K, and effective masses, which reflect the intricate interplay between orbital character, interlayer hybridization, and moiré potential modulation. By comparing them with theoretical models, we provide direct spectroscopic insights into the band structure evolution in this system. Furthermore, we investigate the doping evolution of electronic structure via two complementary approaches: electrostatic gating and surface potassium dosing. These methods enable direct access to both valence and conduction bands, confirming a direct band gap in tMoTe$_2$. By revealing and tuning the momentum-resolved electronic structure across twist angles, we lay the groundwork for advancing both the theoretical understanding and experimental engineering of correlation- and topology-driven phases in twisted TMD moiré materials.

*Contact author: yjdeng@stanford.edu

†Contact author: zxshen@stanford.edu

## II. RESULTS

A schematic of a typical tMoTe$_2$ device used in our ARPES measurements is shown in Fig. 1(a). From top to bottom, the device consists of a graphene layer, tMoTe$_2$ bilayers, a hexagonal boron nitride (hBN) dielectric layer, and a graphite back gate. The top graphene layer, connected to a metal pad, serves both as a protective layer and a reliable electrical ground during measurements. The back gate allows for tuning of the gate voltage ($V_g$) to control doping in the tMoTe$_2$. The entire structure is fabricated on a SiO$_2$/Si substrate. In one of the samples [with twist angle of 3.93°, data included in Fig. 3], the tMoTe$_2$ is encapsulated by graphene on one half and hBN on the other, enabling a direct comparison of the band structure under different dielectric screening environments. ARPES measurements on three graphene-encapsulated areas [Fig. S1(a)-(c) in the Supplemental Material (SM) [24]] and one hBN-encapsulated area [Fig. S1(d) in SM [24]] show comparable band structure parameters [Fig. S1(e) in the SM [24]], indicating good spatial uniformity and suggesting that encapsulation with graphene or hBN has minimal impact on the electronic structure at this twist angle, despite a small rigid band shift [25].

An optical image of a representative sample, illustrating the device geometry, is shown in Fig. 1(b) (left panel). During ARPES measurements, the integrated photoemission map near the Fermi level ($E_F$) clearly resolves sharp features from electrical contact and graphite, locating the sample [shown in Fig. 1(b), middle panel]. Samples prepared using standard pre-measurement treatments, including immersion in organic solvents and annealing in ultra-high vacuum, already show clear photoemission signal from graphene. However, a critical step for acquiring high-quality spectra from the underlying tMoTe$_2$ is the removal of interface residue using an atomic force microscope (AFM) tip in contact mode [details can be found in the Appendix and Fig. S2 in the SM [24]]. This process significantly enhances the ARPES spectra in the AFM-cleaned region [marked by the white dashed line in Fig. 1(b)]. Both photoemission intensity maps integrated near $E_F$ and the Te core level [Fig. 1(b), middle and right panel] reveal notably stronger intensity in this region, which is the primary area used for our measurements.

The twist angles were determined either from the moiré wavelengths $a_M$ observed using piezo-force microscopy (PFM) [Fig. 1(e), see also Appendix] both prior to and after the ARPES measurements, or from the moiré wavevectors $G_M$ extracted from the constant energy contours near the graphene K point [Fig. 1(d), see also Appendix] during the measurements. Figure 1(c) presents a schematic of the Brillouin zones for tMoTe$_2$ with graphene. The relative angle between tMoTe$_2$ and graphene varies across samples, as the two were not intentionally aligned during fabrication. Moiré minibands are theoretically anticipated to emerge near the K points of MoTe$_2$ (refs. [21,26]).

Figure 2 presents the ARPES spectra of pristine monolayer and bilayer 2H-MoTe$_2$, along with tMoTe$_2$ at various twist angles. The upper row shows the raw ARPES spectra along the Γ-K high-symmetry direction, while the lower row displays the corresponding second-derivative spectra with respect to energy, enhancing the visibility of subtle features. For visual comparison, the energy scales are referenced to the VBM at the K point ($E - E_{VBM}$), compensating for sample-dependent Fermi level pinning.

Before diving into the band structure of tMoTe$_2$, we first examine the band structure of monolayer and bilayer 2H-MoTe$_2$ (corresponding to a 60° twist angle), shown in Fig. 2(f) and Fig. 2(a), as benchmarks for understanding the modifications introduced by moiré effects. 2H bilayer sample was exfoliated and measured in situ under ultra-high vacuum on gold substrate, whereas the monolayer MoTe$_2$ was measured in situ on a twisted device. The topmost valence bands at the K point primarily originate from Mo $d_{xy}$ and $d_{x^2-y^2}$ orbitals, with minor contributions from Te $p_x$ and $p_x$ orbitals. In contrast, the topmost valence band at the Γ point arises from the hybridization of Mo $d_{z^2}$ and Te $p_z$ orbitals [27,28]. These distinct orbital characters with different symmetries lead to different manifestations of spin-orbit coupling at K and Γ, ultimately leading to contrasting electronic structures at the two valleys. At the K point, broken inversion symmetry and strong spin-orbit coupling result in two spin-split bands [Fig. 2(f)]. In the H-stacked bilayer, this splitting persists; however, the restored inversion symmetry enforces overall spin degeneracy, rendering each pair of bands spin-degenerate [Fig. 2(a)]. At the Γ point, monolayer exhibits a nearly flat dispersion [Fig. 2(f)], whereas in the bilayer, strong interlayer hybridization of orbitals with out-of-plane character leads to a clear band splitting [Fig. 2(a)] [28,29].

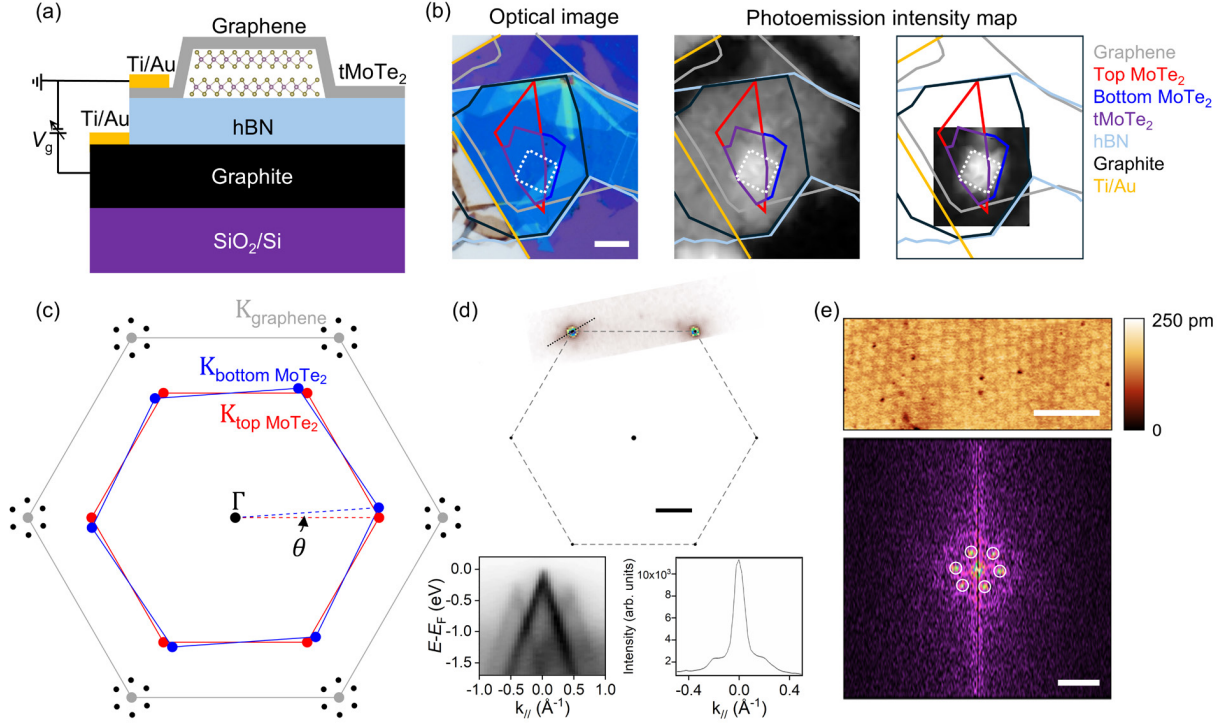

FIG. 1. Device structure and characterization of tMoTe$_2$ heterostructure. (a) Schematic of a tMoTe$_2$ device designed for ARPES measurements. (b) Optical image (left), integrated photoemission intensity map near the Femi level (middle), and near the Te 4d core level (right) of the same region of a tMoTe$_2$ device (twist angle: 1.98°). The AFM-cleaned region is outlined by a white dashed line. Scale bar, 10 μm. (c) Schematic of the Brillouin zones of tMoTe$_2$ with graphene on top (graphene: grey, top MoTe$_2$: red, and bottom MoTe$_2$: blue). (d) Twist angle determination by ARPES constant-energy contour measurements. Top panel: Constant-energy contours near the graphene K points (K$_{graphene}$), showing multiple replica Dirac cones separated by the moiré wavevectors of tMoTe$_2$. Scale bar, 0.5 Å$^{-1}$. Bottom-left panel: Band dispersion along the black dashed line in the top panel. Bottom-right panel: Momentum distribution curve (MDC) at the Dirac points, extracted from the dispersion in the bottom-left panel. The extracted moiré wavevector $G_M$ is 0.21 Å$^{-1}$, corresponding to a twist angle of 5.89°. (e) Twist angle determination by PFM measurements. Top panel: PFM amplitude image of graphene-protected tMoTe$_2$. Scale bar, 50 nm. Bottom panel: Fast Fourier Transform (FFT) analysis of top panel. Scale bar, 0.2 nm$^{-1}$. The extracted moiré wavelength $a_M$ is 10.36 nm, corresponding to a twist angle of 1.98°.

Now we turn to the band structure of tMoTe$_2$ with varying twist angles [Figs. 2(b)-(e), 2(h)-(k) and Fig. S3 in the SM [24]]. In tMoTe$_2$, the valence band structure resembles that of the 2H bilayer, with distinct band splitting at both the K and Γ points. Across all configurations—monolayer, 2H bilayer, and twisted bilayer—the global VBM is always located at the K point, allowing moiré states to be energetically isolated in twisted bilayers. In twisted bilayers, the energy separation between VBM at K and Γ becomes more pronounced compared to the 2H bilayer configuration [Fig. 2(a) and 2 (g)], with VBM at Γ point further lowered as the twist angle increases [Fig. 2(b)-(e) and 2(h)-(k)]. Additionally, a systematic flattening near the valance band top at the Γ point is observed with increasing twist angles. This trend indicates a suppression of interlayer hybridization as the system approaches the monolayer limit, resulting in a simplified band structure characteristic of fully decoupled layers [Fig. 2(f) and 2(l)].

While the qualitative differences in band structure already reveal twist-angle-dependent band reconstructions, a deeper understanding requires a more systematic and quantitative analysis. In particular, variations in Fermi level pinning across different twist angles complicate direct comparisons of band structure. We therefore extracted three key valence-band parameters: energy splitting between the topmost two valence bands at the Γ and K points (denoted as $\Delta_\Gamma$ and $\Delta_K$, respectively), and the energy difference between VBM at the K and Γ points ($\Delta_{K-\Gamma}$). Their evolution with twist angle, shown in Fig. 3, provides further insight into the interplay between interlayer coupling and moiré modulation.

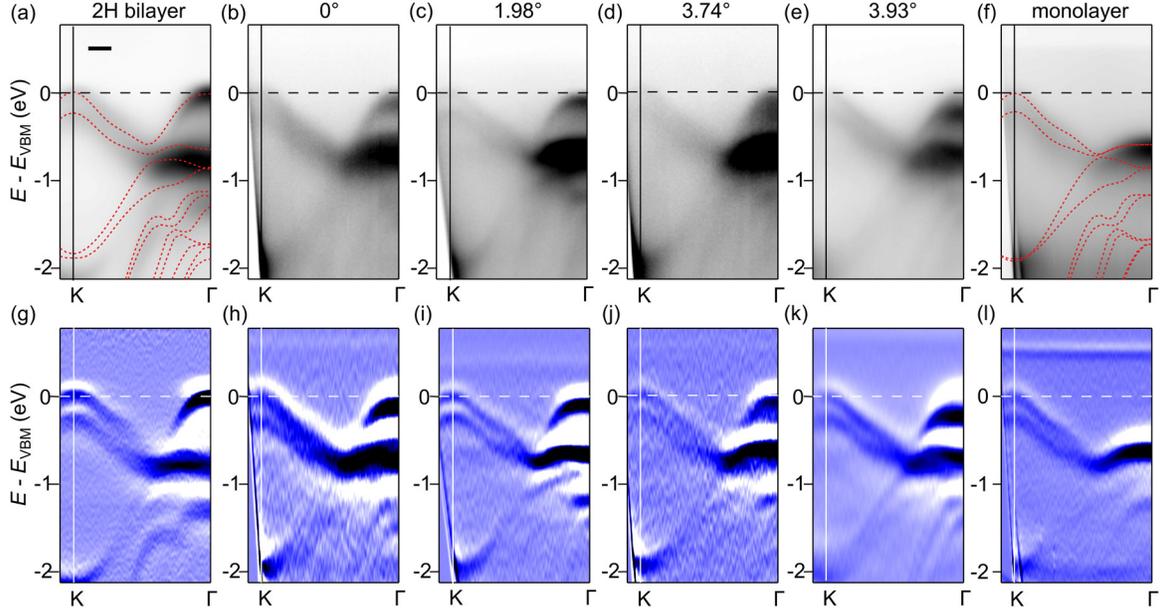

FIG. 2. Electronic band structure of 2H bilayer, twisted bilayer and monolayer MoTe$_2$. (a)-(f) ARPES band dispersion along the Γ−K high-symmetry direction in 2H bilayer, twisted bilayers at representative twist angles, and monolayer MoTe$_2$. Overlaid red curves show DFT-calculated band structures using vdW-DF2 functional, energy-shifted to align with experimental features. Scale bar, 0.2 Å$^{-1}$. (g)-(l) Corresponding second derivative spectra of a-f with respect to energy. To enable direct comparison, the VBM at the K point is set to zero energy (black dashed lines in (a)-(f), white dashed lines in (g)-(l)).

As shown in Fig. 3(a), $\Delta_\Gamma$ decreases monotonically with increasing twist angle. This trend reflects the weakening interlayer hybridization between out-of-plane orbitals (Mo $d_{z^2}$ and Te $p_z$), driven by reduced atomic registry and diminished orbital overlap. At small twist angles, strong lattice relaxation promotes the formation of triangular AB/BA stacking domains with reduced interlayer spacing, leading to strong interlayer coupling. Since no moiré minibands form at the Γ point, the observed $\Delta_\Gamma$ provides a direct measure of twist-angle-dependent interlayer coupling.

In contrast, Figure 3(b) shows that $\Delta_K$ remains relatively constant across twist angles, exhibiting only minor modulations before decreasing toward the monolayer limit. This weak dependence stems from the nature of the states living at the K point, which originate primarily from in-plane orbitals (Mo $d_{xy}$, $d_{x^2-y^2}$ and Te $p_x$, $p_x$) and are dominated by spin-orbit coupling within individual layers, making them relatively insensitive to changes in vertical interlayer tunneling. The slightly larger $\Delta_K$ observed in bilayers than in monolayers is attributed to enhanced interlayer coupling.

As a result, $\Delta_{K-\Gamma}$ increases systematically with twist angle, as shown in Fig. 3(c). This behavior arises from the contrasting evolution of $\Delta_\Gamma$ and $\Delta_K$: while VBM at the Γ point shifts downward due to suppressed interlayer hybridization, VBM at the K point remains nearly unchanged owing to its predominantly in-plane orbital character and weaker sensitivity to interlayer coupling. The resulting increase in energy separation underscores the differing responses of the two valleys to twist-induced interlayer effects.

We performed DFT calculations to capture the evolution of $\Delta_{K-\Gamma}$ with twist angle, incorporating lattice relaxation and spin-orbit coupling. To assess the reliability of different approaches, we tested various van der Waals (vdW) functionals. While atom-wise vdW functionals such as DFT-D2, DFT-D3, and dDsC failed to reproduce even the correct trend for untwisted bilayers, non-local van der Waals functionals including vdW-DF2, vdW-DF3-opt2, SCAN+rVV10, and r$^2$SCAN+rVV10 [see Fig. S5 in the SM [24], see also Appendix] qualitatively reproduce the overall trend of increasing $\Delta_{K-\Gamma}$ with twist angle. Among them, the vdW-DF2 produces results closest to the experiment for untwisted systems, which we adopt here [Fig. 3(c) and 3(f)] and for the rest of the paper. The corresponding calculated band structures are shown in Fig. S6 of the SM [24]. However, the calculated $\Delta_{K-\Gamma}$ is expected to deviate further from the experimental trend at twist angles below 5°, based on a reasonable extrapolation of the DFT results [Fig. 3(f)]. This discrepancy suggests that mean-field approximations inherent in standard DFT are insufficient to fully capture the electronic structure of tMoTe$_2$, possibly due to enhanced electron-electron correlation effects beyond the DFT level.

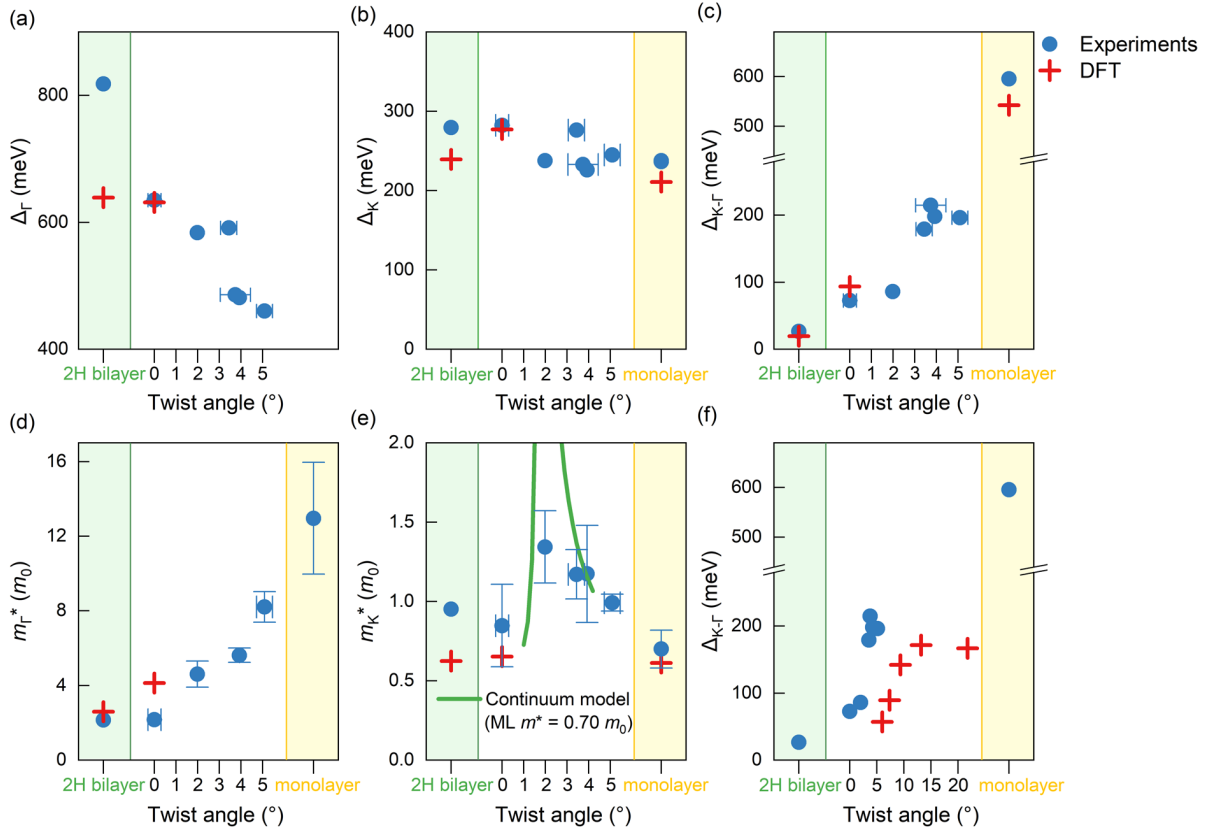

FIG. 3. Evolution of band structure parameters and effective masses in 2H bilayer, twisted bilayers, and monolayer MoTe$_2$. (a)-(c) Twist-angle dependence of key valence band parameters: $\Delta_\Gamma$ (a), $\Delta_K$ (b) and $\Delta_{K\text{-}\Gamma}$ (c). For $\Delta_{K\text{-}\Gamma}$ we compare the $\Gamma$-point VBM in both monolayer and bilayers, but bilayer splitting makes it a different band, so the monolayer value is shown as a reference. (d) and (e) Twist-angle dependence of $m_\Gamma^*$ and $m_K^*$. The continuum model (green line) uses a monolayer effective mass of 0.70 $m_0$ (obtained from experimental fitting) and interlayer coupling parameters $V = -11.2$ meV and $\omega = 13.3$ meV as inputs. Both experiments and theory show a peak near 2°, indicating maximal band flattening at this critical angle. Error bars on the effective mass represent fitting uncertainties. (f) $\Delta_{K\text{-}\Gamma}$ plotted over an extended twist-angle range, complementing panel (c). In all panels, experimental data are shown as blue dots; DFT results using the vdW-DF2 functional are shown as red crosses. The DFT value at 0° twist angle is obtained from a calculation of a rhombohedral-stacked bilayer. Green and yellow shaded regions represent the 2H bilayer (60°) and monolayer limits, respectively. Twist angle uncertainties are derived from PFM or ARPES measurements. Error bars smaller than the marker size are not shown.

To further assess the influence of interlayer coupling and moiré modulation on band flattening, we extracted hole effective masses at the $\Gamma$ and K points ($m_\Gamma^*$ and $m_K^*$) from ARPES measurements across a range of twist angles. The effective masses were obtained by fitting the curvature of the topmost valence band near the $\Gamma$ and K points within a 0.2 Å$^{-1}$ momentum window using the parabolic approximation $E(k) = \hbar^2 k^2 / 2m_i^*$, where $i = \Gamma$ or K. The extracted values are shown in Fig. 3(d) and 3(e).

As shown in Fig. 3(d), $m_\Gamma^*$ at small twist angles (~0°) is close to the 2H bilayer value, which serves a reference. With increasing twist angle, $m_\Gamma^*$ rises significantly, reaching 8.2 $m_0$ near 5°, where $m_0$ is the free electron mass, and approaching the heavy monolayer limit value of ~13 $m_0$. This trend reflects a reduction in both interlayer hybridization and lattice relaxation with increasing twist angle, leading to flatter valence bands at $\Gamma$. In contrast, $m_K^*$ remains relatively constant across most twist angles, consistent with the dominant in-plane orbital character of the K-point states and their weak sensitivity to vertical interlayer coupling. However, a distinct peak appears near 2°, where $m_K^*$ reaches a local maximum. This non-monotonic behavior suggests maximum band flattening at this critical angle. Although individual moiré minibands are not resolved in the spectra, the effective mass reflects an average over these dispersions. Such a feature is indicative of emerging interaction-driven phenomena in tMoTe$_2$.

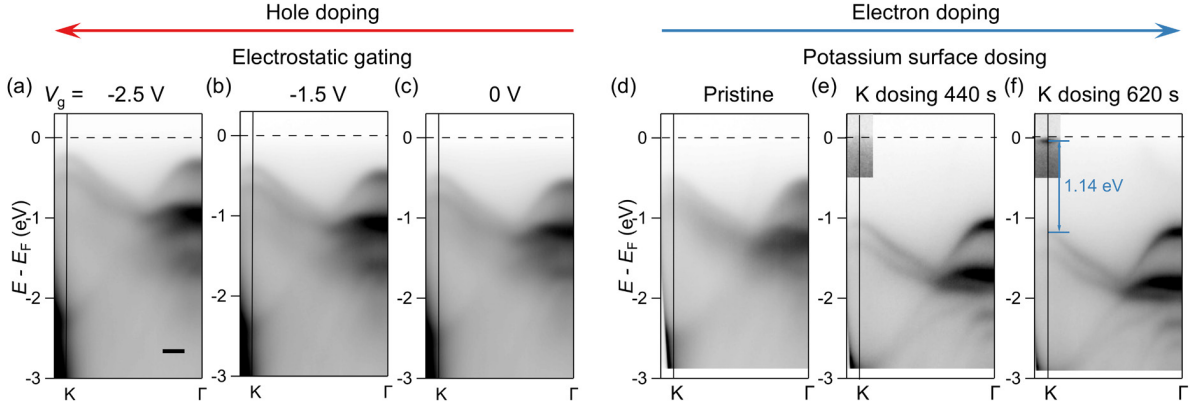

FIG. 4. Doping evolution of the electronic structure in tMoTe$_2$. (a)-(c) ARPES band dispersion along the Γ−K high-symmetry direction of 1.98° tMoTe$_2$ under different back-gate voltages ($V_g$: –2.5 V, –1.5 V and 0 V), showing systematic upward shifts of the valence band edge with increasing hole doping. Scale bar, 0.2 Å$^{-1}$. (d)-(f) ARPES band dispersion along the Γ−K high-symmetry direction of 3.44° tMoTe$_2$ before (d) and after in-situ potassium surface dosing for 440 s (e) and 620 s (f), illustrating progressive downward band shifts corresponding to electron doping. In e and f, the data near the K point close to $E_F$ is shown with an adjusted color scale to enhance contrast. While no CBM is visible in (e), it becomes clearly resolved in (f), revealing a direct band gap of 1.14 eV.

Having established how twist angle modulates the valence band structure and hole effective masses in tMoTe$_2$, we next explore how external doping, implemented through electrostatic gating and surface potassium dosing, enables tunability of the Fermi level and access to the conduction band.

In a 1.98° tMoTe$_2$ device, we shifted the Fermi level toward the hole-doped regime by applying negative gate voltages. As shown in Fig. 4(a)-(c), increasingly negative $V_g$ shifts the K-point VBM upward, confirming progressive hole doping, while the underlying band dispersions and key band structure parameters remain essentially unchanged up to $V_g = -2.5$ V [Fig. S9(e) in the SM [24]]. We also gated the same device towards the electron side, up to a conservative +2.5 V to avoid sample damage, the conduction band was not observed [see Fig. S9(d) and 9(i) in the SM [24]]. This may results from the limited bias as well as the reduced gating efficiency in the presence of the graphene capping layer [30]. Additionally, the spectra become substantially broadened under positive gating, likely due to local charge and potential inhomogeneity, consistent with previous observations under similar gating conditions [31].

In a separate 3.44° tMoTe$_2$ device, efficient electron-side doping was achieved via in-situ potassium dosing [Fig. 4(d)-(f)]. As dosing proceeds, the Fermi level shifts upward, and the spectral features sharpen noticeably, which could be attributed to enhanced screening of the Coulomb interaction and a subsequent reduction in electron-electron scattering. After 620 seconds of dosing, the conduction band minimum (CBM) became clearly resolvable, revealing a direct band gap of 1.14 eV [Fig. 4(f)]. Importantly, both monolayer and 2H bilayer MoTe$_2$ are known to host direct band gaps [32,33], supporting that the direct band gap observed in tMoTe$_2$ reflects the intrinsic band structure of MoTe$_2$. Our DFT calculations [Fig. S6 and S10 in the SM [24]], conducted across a broad range of twist angles, consistently show that tMoTe$_2$ remains a direct-gap semiconductor with a nearly constant band gap, in excellent agreement with our experimental value.

These results demonstrate how electrostatic gating and potassium dosing, as complementary modulation methods, enable carrier-density control in moiré materials, highlighting their potential for device-level engineering.

## III. DISCUSSION AND CONCLUSION

Although moiré minibands are expected to appear near the K valley in tMoTe$_2$ [21,26], our ARPES measurements did not observe such features. Photon energy-dependent measurements across the K valley [Fig. S4 in the SM [24]] show no evidence of minibands, suggesting that matrix element effects are unlikely to account for their absence. Nevertheless, the presence of a moiré potential in our tMoTe$_2$ samples, indicated by either replica bands near the K valley of graphene or moiré patterns detected by PFM, suggests that the lack of observable minibands stems from other factors. The most likely limiting factor is quasiparticle lifetime broadening at measurement temperature of 20 K, which, combined with the small expected moiré band gap [26] relative to our experimental resolution, makes it exceedingly difficult to resolve such subtle features. In addition, the moiré potential at the K point is inherently weaker, less than one-fifth of that at the Γ point according to continuum model calculations [26,34,35], due to the weak interlayer coupling of the relevant orbitals (Mo $d_{xy}$, $d_{x^2-y^2}$ and Te $p_x$, $p_x$) at K.

To gain further insight into the origin and implications of the observed band flattening, we now discuss the theoretical modeling results in light of the experimental trends. For small twist angles, direct DFT calculations become practically inaccessible due to the large moiré supercells, making the continuum model essential for capturing the band structure evolution. Our continuum model calculations, using the experimentally measured monolayer effective mass ($0.70\,m_0$) as input, predict maximal band flattening near 2°, in good agreement with our experimental data [Fig. 3(e), see also Appendix and Fig. S7 and S8 in the SM [24]), and consistent with earlier continuum model studies [21,23]. It is worth noting that while the continuum model computes the effective mass of the topmost moiré miniband, the experimentally extracted values reflect an average over unresolved minibands. Nevertheless, the model captures the overall trend and identifies the critical twist angle. Although lattice relaxation is not explicitly included in our continuum Hamiltonian, its effects are effectively incorporated through DFT-informed parameters, enabling quantitatively accurate modeling (see Appendix). Finally, we note DFT calculations underestimate the effective masses for monolayer, 2H bilayer and 0° tMoTe₂ at the K point, again underscoring the necessity of beyond-DFT treatments to capture correlation effects in this system.

In conclusion, our combined experimental and theoretical investigation of tMoTe₂ reveals a systematic evolution of the valence band structure and hole effective mass with twist angle. ARPES measurements uncover pronounced band reconstruction at the Γ point and non-monotonic band flattening at the K point, identifying ~2° as a magic angle, consistent with continuum model calculations. Additionally, we establish electrostatic gating and surface dosing as effective methods for modulating the Fermi level and accessing conduction states. These findings position twisted MoTe₂ as a tunable platform for engineering correlated electronic states.

Looking forward, our results motivate future studies at lower temperatures and higher energy resolution, as well as gate-tunable ARPES experiments, to resolve the elusive moiré minibands and explore interaction-driven phases near the magic angle.

## APPENDIX: METHODS
### 1. Sample fabrication

tMoTe₂ devices were fabricated inside a glove box using two methods. Most samples were prepared using Method 1, while the 1.98° sample was prepared using Method 2. hBN, graphene and MoTe₂ flakes were prepared by mechanical exfoliation onto O₂ plasma-treated SiO₂/Si substrates.

Method 1:

The tMoTe₂ device was fabricated following the approach of ref. [36] with minor modifications. The hBN, which eventually serves as the bottom gate dielectric, was first picked up using a polycarbonate (PC) film on a dome-shaped polydimethylsiloxane (PDMS) block. Then we employed the tear-and-stack method [37] to sequentially assemble two halves of exfoliated MoTe₂ at the desired twist angle, followed by a graphene flake, all at 90 °C. The hBN/tMoTe₂/graphene stack was transferred onto a gold-coated PDMS block by melting the PC at 180 °C. The stack was cleaned by sequential rinsing in N-methyl-2-pyrrolidone and isopropanol, blow with a nitrogen gas gun and then transferred onto a graphite flake on a SiO₂/Si substrate or a highly conductive silicon substrate. Au/Ti (20 nm/3 nm) electrodes for grounding or gating, were deposited through stencil masks inside a glove box.

Method 2:

The hBN bottom gate dielectric layer was pre-transferred onto the graphite gate using a PC film on a PDMS block. Pt/Ti (30 nm/5 nm) electrodes were pre-patterned on the substrate via electron-beam lithography, with a small electrode contacting the graphite gate and a larger grounded electrode surrounding the sample to minimize electrostatic distortion during gating. A graphene flake and two halves of exfoliated tMoTe₂ were sequentially picked up with PC and transferred onto the hBN bottom gate by melting the PC at 180 °C, followed by rinsing in chloroform and drying with a nitrogen gas gun.

Additionally, the 2H bilayer sample was prepared by in-situ exfoliation of bulk single crystals onto gold-coated SiO₂/Si substrates.

### 2. AFM measurements

Mechanical cleaning of polymer residues at sample interfaces was performed using an Oxford Instruments Asylum Research Cypher AFM in contact mode with BudgetSensors Tap300-G tips (force constant: 40 N/m, tip radius: < 10 nm). The setpoint voltage was adjusted between 0-0.3 V to accommodate tip wear during scanning, with a scan rate of 0.1-0.15 Hz over a 10 μm × 10 μm scan area.

Subsequent twist-angle determination was conducted via vector PFM using Ir/Pt-coated conductive tips (Nanosensor PPP-EFM, force constant: 2.8 N/m, tip radius: < 25 nm). An AC bias of 1-3 V was applied during scanning. Typical resonance frequencies were 300-350 kHz for vertical and 750-850 kHz for lateral PFM. The setpoint voltage was adjusted between 0-0.4 V during the scan. FFT analysis of the PFM amplitude image yielded the moiré wavelength $a_M$, allowing calculation of the twist angle $\theta$ using $\theta = 2\arcsin(a_0/2a_M)$, where the MoTe$_2$ lattice constant $a_0$ is taken as 0.355 nm.

### 3. ARPES measurements

ARPES measurements of tMoTe$_2$ devices were conducted at Beamline 7.0.2 (MAESTRO) of the Advanced Light Source (ALS) at Lawrence Berkeley National Laboratory. Samples were annealed at 200 °C in vacuum for 1.5 hours and measured at ~ 20 K under a pressure below $3 \times 10^{-11}$ torr. The photon energy was 58 eV for all data shown in the main text. Spectra were collected using a Scienta R4000 analyzer with deflectors, and the photon beam was focused to a 1-2 μm spot using a capillary mirror. The energy and angular resolutions were 10-20 meV and 0.2°, respectively.

ARPES measurements of 2H bilayer and monolayer MoTe$_2$ were performed at Beamline 5-2 of the Stanford Synchrotron Radiation Lightsource (SSRL) at SLAC National Accelerator Laboratory. The samples were exfoliated in situ at 200 K and measured at 7 K under a pressure below $2 \times 10^{-11}$ torr. Data were collected using a Scienta DA 30 analyzer with deflectors with 58 eV photons and ~ 10 meV energy resolution. The beam spot size was ~ 5 μm × 32 μm.

The moiré potential of tMoTe$_2$ modulates the Dirac electrons in the top graphene layer, producing replica Dirac bands arranged hexagonally around the $K_G$ point [Fig. 1(c) [38]]. These replicas are separated by the moiré wavevector $\boldsymbol{G}_M$ of tMoTe$_2$, whose magnitude was extracted from an MDC taken across the Dirac points along the $\boldsymbol{G}_M$ direction [Fig. 1(d), bottom-right panel]. The twist angle $\theta$ was then determined using the relation $\theta = 2\arcsin\left(\frac{\sqrt{3}G_M \cdot a_0}{8\pi}\right)$.

### 4. DFT calculations

The DFT calculations were performed with the Vienna Ab initio Simulation Package (VASP) [39,40]. Van der Waals (vdW) forces are weak, long-ranged interactions that significantly modify the mechanical properties and band structures of layered materials. We incorporated vdW forces through non-local vdW functionals that do not rely on empirical fitting or predefined atom-pairwise dispersion coefficients like the DFT-D2 functional [41]. Specifically, we compared four different non-local vdW functionals: vdW-DF2 [42], vdW-DF3-opt2 [43], SCAN+rVV10 [44], r$^2$SCAN+rVV10 [45].

To obtain the DFT band structures, we included both structural relaxation and spin-orbit coupling to best match the experimental observations. We followed the following workflow. First, we relaxed the atomic positions with fixed lattice parameter of 3.55 Å without spin-orbit coupling. With the fully relaxed structure, we performed self-consistent calculations that include spin-orbit coupling. Finally, the band structure was calculated along high-symmetry line cut using the self-consistent wavefunctions. For untwisted systems, we relaxed the structure to electronic convergence of $1 \times 10^{-6}$ eV/Å and force convergence of $1 \times 10^{-4}$ eV/Å. We used k-point sampling of $6 \times 6 \times 1$ for the self-consistent calculations. For twisted supercell calculations, we first constructed the supercell according to the following relation:
$$\theta = \text{acos}\left[\frac{N^2 + 4MN + M^2}{2(N^2 + NM + M^2)}\right],$$
where $M$ and $N$ are two integers. We chose $M = N + 1$ so that the periodic supercell and moiré cell are the same. We used electronic convergence of $1 \times 10^{-4}$ eV/Å and force convergence of $1 \times 10^{-3}$ eV/Å, as well as k-point sampling of $5 \times 5 \times 1$ for the self-consistent calculations.

### 5. Continuum model calculations

To investigate the moiré minibands of tMoTe$_2$, we employed a K-valley continuum model constructed for hole states near the K and K' points of the Brillouin zone, where spin-orbit coupling leads to a single spin-valley locked band per layer. The twist introduces a moiré superlattice potential and modifies interlayer tunneling, both incorporated in the model. This approach is appropriate for hole-doped samples where the Fermi level lies near the valence band top at the K point and the Γ-point states are far below in energy.

The continuum model Hamiltonian was constructed in the basis of layer pseudospin. The coordinate origin was set midway along the rotation axis of the two layers. We focused on the spin-up component ($\mathcal{H}_\uparrow$):
$$\mathcal{H}_\uparrow = \begin{pmatrix} \frac{\hbar^2(-i\nabla_{-\kappa_+})^2}{2m} + V_1(\boldsymbol{r}) & t(\boldsymbol{r}) \\ t^\dagger(\boldsymbol{r}) & \frac{\hbar^2(-i\nabla_{-\kappa_-})^2}{2m} + V_2(\boldsymbol{r}) \end{pmatrix},$$

while the spin-down component ($\mathcal{H}_\downarrow$) is the time reversal conjugate.

The corners of the moiré Brillouin zone are $\kappa_\pm = \frac{4\pi}{3a_M}\left(\frac{\sqrt{3}}{2}, \pm\frac{1}{2}\right)$. $V_l(\mathbf{r})$ and $t(\mathbf{r})$ represent the intralayer moiré potential and interlayer tunneling, respectively, for layer index $l$. Incorporating $D_3$ symmetry and retaining only the lowest Fourier harmonics, we expressed these terms as:
$$V_l(\mathbf{r}) = -2V \sum_{i=1,3,5} \cos(\mathbf{g}_i + \phi_l),$$
$$t(\mathbf{r}) = \omega\left(1 + e^{-i\mathbf{g}_2 \cdot \mathbf{r}} + e^{-i\mathbf{g}_3 \cdot \mathbf{r}}\right).$$

Here, the moiré reciprocal lattice vectors are $\mathbf{g}_i = \frac{4\pi}{\sqrt{3}a_M}\left(\cos\left(\frac{\pi(i-1)}{3}\right), \sin\left(\frac{\pi(i-1)}{3}\right)\right)$ for $i = 1, \dots, 6$, and $\phi_1 = -\phi_2 = \phi$.

We retained only the valence band sector of the model, consistent with the experimentally hole-doped regime. The resulting minibands were obtained by diagonalizing the continuum Hamiltonian in a plane-wave basis. Following the standard convention, we defined the top (second) moiré valence band as the highest (second-highest) eigenvalue at a given $\mathbf{k}$ point within the moiré Brillouin zone.

While the Hamiltonian does not include lattice relaxation explicitly, its effects are effectively accounted for through the renormalized moiré potential $V_l(\mathbf{r})$ and interlayer tunneling $t(\mathbf{r})$, whose amplitudes $V$ an $\omega$ are extracted from DFT calculations that incorporate atomic relaxation. Lattice relaxation primarily suppresses $V$ and $\omega$, introduces higher harmonics, and induces local gap corrections that further separate K- and Γ-derived bands. These effects together make the K-valley model a quantitatively accurate description of the valence band physics in tMoTe$_2$.

Finally, to compare with experiment, we extracted the effective mass from a parabolic fit to the top moiré valence band near $k = 0$, using $m^* = \hbar^2/(\partial^2 E(k)/\partial k^2)$. The use of an isotropic effective mass is justified by the six-fold rotational symmetry of the emergent triangular moiré lattice, which enforces equal curvatures along the Γ-K and Γ-M directions to leading order, consistent with the near-circular constant energy contours observed in this region.

## ACKNOWLEDGEMENTS

We thank K.-J. Xu, C. Lin, and E. Corbae for assistance with ARPES measurements; G. Zaborski, K. Crust, and A. Khandelwal for support with PFM measurements; M. Pendharkar and S. Tran for help with torsional force microscopy measurements; P. Nguyen, H. Park, N. Wang and Y. Yu for assistance with sample fabrication; and A. Reddy for providing code used to build the initial continuum model.

Y.D., P.M., M.H., D.L., and Z.-X.S. acknowledge the support of the U.S. Department of Energy, Office of Science, Office of Basic Energy Sciences, Division of Material Sciences and Engineering, under Contract No. DE-AC02-76SF00515. The sample fabrication and characterization at U. Washington (W.H. and X.X.) are supported by Department of Energy, Basic Energy Sciences, Materials Sciences and Engineering Division (DE-SC0012509). Y.D. acknowledges partial support from a Stanford Q-FARM Bloch Postdoctoral Fellowship. Z.Z. acknowledges support from a Stanford Science fellowship. T.Z. acknowledges support from the MIT Dean of Science Graduate Student Fellowship. L.F. acknowledges partial support from a Simons Investigator Award from the Simons Foundation. P.M. acknowledges support from a Stanford Energy Postdoctoral Fellowship through contributions from the Dai and Li Family Stanford Sustainability Postdoctoral Fellow Program Fund, Precourt Institute for Energy, Bits & Watts Initiative, StorageX Initiative, and TomKat Center for Sustainable Energy. K.W. and T.T. acknowledge support from the JSPS KAKENHI (Grant Numbers 21H05233 and 23H02052), the CREST (JPMJCR24A5), JST and World Premier International Research Center Initiative (WPI), MEXT, Japan. Use of the Advanced Light Source, Lawrence Berkeley National Laboratory, is supported by the U.S. Department of Energy, Office of Science under Contract No. DE-AC02-05CH11231. Use of the Stanford Synchrotron Radiation Lightsource, SLAC National Accelerator Laboratory, is supported by the U.S. Department of Energy, Office of Science, Office of Basic Energy Sciences under Contract No. DE-AC02-76SF00515. Use of the Stanford Nano Shared Facilities (SNSF), is supported by the National Science Foundation under award ECCS-2026822.

## AUTHOR CONTRIBUTIONS

Y.D. and Z.-X.S. conceived and supervised the project. Y.D. and W.H. fabricated the samples. Y.D. and P.M. conducted ARPES measurements at ALS beamline 7, while Y.D. performed additional ARPES measurements at SSRL beamline 5. Y.D. analyzed the ARPES data. T.T. and K.W. synthesized the high-quality hBN crystals. C.J., A.B., and E.R. maintained ALS beamline 7, while M.H. and D.L. maintained SSRL beamline 5. Z.Z. and T.P.D. provided DFT calculations. T.Z. and L.F. provided continuum model calculations. Y.D. wrote the paper with contributions from all authors.